\documentclass[10pt]{article}

\usepackage{epsfig}
\usepackage{amsmath}
\usepackage{natbib}
\usepackage{tabularx}
\usepackage{subfig}
\usepackage{float}

\begin{document}

\title{Communicating extraterrestrial intelligence (CETI) interaction models based on the Drake Equation}
\author{Reginald D. Smith \\ Ronin Institute \\ 127 Haddon Pl \\ Montclair, New Jersey 07043 \\  \\Supreme Vinegar LLC \\ 3430 Progress Dr. Suite D \\ Bensalem, PA 19020, USA
\\ \texttt{rsmith@supremevinegar.com}}

\maketitle

\begin{abstract}
The Drake Equation has proven fertile ground for speculation about the abundance, or lack thereof, of communicating extraterrestrial intelligences (CETIs) for decades. It has been augmented by subsequent authors to include random variables in order to understand its probabilistic behavior. However, in most cases, the emergence and lifetime of CETIs are assumed to be independent of each other. In this paper,  we will derive several expressions that can demonstrate how CETIs may relate to each other in technological age as well as how the dynamics of the concurrent CETI population change under basic models of interaction, such as the Allee Effect. By defining interaction as the change in the expected communication lifetime with respect to the density of CETI in a region of space, we can use models and simulation to understand how the CETI density can promote or inhibit the longevity and overall population of interstellar technological civilizations.
\end{abstract}

\section{Introduction}

The Drake equation has been a useful tool for thought experiments about the presence (or lack thereof) of intelligent, communicating extraterrestrial intelligences (hereafter ``CETI'') for over fifty years. In recent years, the Drake equation, long viewed as a deterministic type relation, has been supplemented with the tools of the probability and statistics of random variables \citep{statdrake1,prantzos1, poisson,prantzos2}. Assuming that the terms in the Drake equation are random variables, this allows us to express the Drake equation results as first moments of a distribution with confidence intervals.

In \citep{poisson}, the Drake equation is refined by defining the arrival rate of communicating civilizations as a Poisson distribution with their lifetime modeled as a random variable. Monte Carlo simulations \citep{forgan1,forgan2,hair} have allowed the approximation of various parameters for the emergence and permanence of CETI depending on underlying models of fundamental factors. The possibility of interaction between earlier, or first, civilizations and their predecessors was broached by \citep{hair} by using Monte Carlo simulations to model the distribution of times of the arrival of a first civilization and later CETI inter-arrival times. One of the results is that there was likely a huge interval between the arrival of the first technological civilization and its successors which theoretically could allow it a long time to leave a legacy that would have a disproportionate influence on subsequent societies.

Many papers have modeled the probabilities behind the likelihood of emergence as well as durability of CETI. Few have gone into similar quantitative detail on the nature of the interactions between CETI.  In this paper we will address the questions of the average time between successive civilizations though unlike \citep{hair} we are looking at just the steady state, not the time between the first one or first few CETI. Second, we will measure the probability that CETIs will be able to communicate at all given their random emergence in a volume of space. Finally, we will derive basic expressions for the interaction of CETIs where the expected lifetime of a CETI depends on the density of concurrent CETIs it can interact with. This imputes a covariance between $N$ and $L$ for concurrent CETI that could speak in general terms about the typical fate of interaction between interstellar technological civilizations.

\section{Expected technological `age' gap}

As stated earlier, the emergence of communicating civilizations can often be approximated as a Poisson distribution. Assume $k$ is the number of CETIs that emerge in a given interval, and $\lambda$ is the average arrival rate for CETIs per unit time. The Poisson distribution of the CETI emergence, $N_b(k,\lambda)$ , is given by

\begin{equation}
N_b(k,\lambda)=\frac{\lambda^k}{k!}e^{-\lambda}
\end{equation}

Note this assumes that the Poisson distribution is homogenous so that $\lambda$ is a constant that does not change over time. In queueing theory, given a homogenous Poisson arrival process, one of the key findings of operations research is Little's Law, that states the average number of items in a queue is equal to the arrival rate times the queue handling time. In \citep{smith} it was shown that previous work by Brian Tung (private communication) at his website Astronomy Corner showed that the Drake Equation was a special case of Little’s Law \citep{little}. Where $N$ is the average number of items in a system, $\lambda$ is the average (stationary) arrival rate and $t$ is the average time an item is in the system, Little's Law is expressed as

\begin{equation}
N=\lambda t
\label{little}
\end{equation}

Given the Drake equation \citep{drake}

\begin{equation}
N = R^{*} \times f_s \times L
\label{drake}
\end{equation}

where $ R^{*}$ is the average production rate for stars `suitable' for planets and eventually intelligent life, $f_s$ is the probability of the emergence of an intelligent and communicating civilization around one such star, and $L$ is the average lifetime of such a CETI, the variable $\lambda=R^{*} \times f_s$ and $t=L$.

Therefore, in its simplest form, the Drake Equation and the average number of concurrent civilizations in a region of space can be seen as analogous to some queueing problems. A related question that can be explored from this type of reasoning is the average difference in ``age'' between communicating civilizations given as the difference in the amount of time each has been broadcasting. This would be based on the average emergence time for a civilization and the contingency that there are at least two existing civilizations at any given time in range of contact.

The average interarrival period for a homogenous Poisson process is given by the inverse of the arrival rate, $1/\lambda$. Likewise, on average, at least two civilizations will exist at the same time if $L \geq 1/\lambda$. Given, equations \ref{little} and \ref{drake}, where $N \geq 2$, the average gap in civilization technological age, $\langle t_{gap} \rangle$ can be approximated by

\begin{equation}
\langle t_{gap} \rangle = \frac{1}{\lambda}=\frac{1}{R^{*} \times f_s}=\frac{L}{N}
\label{agediff}
\end{equation}

Therefore, CETIs are most likely to have emerged relatively recently to each other, and possibly have a remotely similar level of technological development, if $L$ is small compared to $N$. In this case, there are many co-existent CETIs of relatively short lifetime. On the other hand, long-lasting CETIs of long $L$ that are fewer in number will more likely have long times between emergence and likely much larger technological gaps. Therefore, this measure could be considered a superficial metric of differential technological advancement though these assumptions beyond the scope of this paper.

\subsection{Existence versus contact}

In \citep{smith}, a model was demonstrated showing that the maximum distance a broadcast from a CETI is expected to have appreciable signal-to-noise, $D$, combines with $L$ to determine requirements for civilization density to hope to achieve any contact. In \citep{prantzos1,prantzos2} a more detailed analysis about the density and possibility of contact between ETIs was made using assumptions of average distances between CETIs assuming they are isotropically distributed throughout the galaxy and using the mathematics of packing to determine the overall density. Here we will explore a slightly different problem of the probability that two co-existing civilizations in a region of space can receive each other’s messages or communicate.

For two CETI to feasibly communicate with each other at least once, the following condition must apply. Namely, that a signal emitted by one CETI should have a reasonable probability of reaching another CETI at most $L/2$ years after emission which should allow a bi-directional communication if there is a prompt reply. A corollary is that the volume of space whose points are $L/2$ light years or less distant from a CETI must have a reasonable probability of $N \geq 2$ given the Drake Equation parameters. Since the Drake Equation is defined over a the entire Milky Way this could cause issues for short lived civilizations if $N$ is small.

To further address the key assumption, a pertinent question is that if a CETI were to send out a signal, given the expanding area of space that signal encounters, how probable is it that the signal reaches another CETI and then how probable is it that the receiving CETI is able to send a response that can be received by the first before its quiescence. From our definition earlier, using the Poisson assumption the emergence rate for CETI is $\lambda=N/L$ for random points emerging in the plane of a circle of radius, $R$, the probability distribution for the shortest distance between any two random points is given by \citep{kendall-moran} (p. 38).

\begin{equation}
\begin{aligned}
f(R)=&2\frac{\lambda}{\pi R_G^2} \pi R \exp{\Bigg(-\frac{\lambda}{\pi R_G^2} \pi R^2 \Bigg)} \\
f(R)=&2\frac{N}{L}\frac{R}{R_G^2} \exp{\Bigg(-\frac{N}{L}\Bigg(\frac{R}{R_G}\Bigg)^2 \Bigg)}
\end{aligned}
\label{eq:randpoints}
\end{equation}

The expression $\frac{N}{L\pi R_G^2}$ is the emergence rate normalized by the surface area of the Milky Way galactic disk simplified as a circle with galactic radius $R_G=50,000$ light-years. Thus the probability of joint communication between CETI is given by calculating the probability of $R$ from zero to $L/2$.

\begin{figure}[H]
\centerline{\includegraphics[height=4in, width=4in]{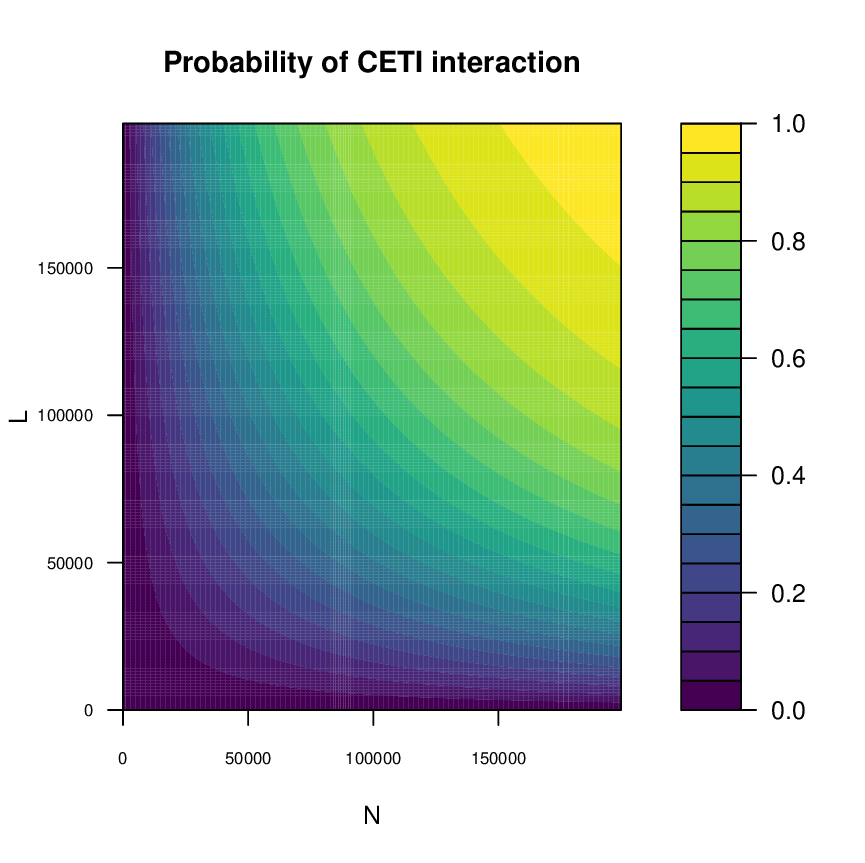}}
\caption{A contour plot of the probability of bi-directional communication between CETI given $N$ and $L$ in the Milky Way. Based on Equation \ref{eq:randpoints}.}
\label{fig:seti-prob}
\end{figure}

In Figure \ref{fig:seti-prob} is a density plot of the probability of two CETI having bi-directional communication based on values of $N$ and $L$.

The probability here should be viewed as a maximum probability given the estimates of CETI emergence rates and lifetime. In short, it is the probability that two CETI can feasibly communicate given the emergence rate for CETI per $L/2$ years per light years cubed. This of course is assuming they will communicate if possible but just because two CETI can communicate doesn't mean they will and it also doesn't mean they find each other even when looking. However, this can provide us with interesting estimates for a lower bound rate for the emergence of CETI.

\section{Interstellar civilization interaction – a definition}

The interaction, or at least awareness, between at least two interstellar technological civilizations is the ostensible goal of all types of SETI. How this could possibly affect the dynamics of a postulated wider interstellar community is the subject of countless works of fiction as well as philosophical and scientific essays. Here we will make a simplifying assumption that any two CETI that meet the previous threshold of being within $L/2$ light-years distance can and will interact. While this is a broad assumption, it will allow us to model and interpret the dynamics of interaction even though in it may not be a given that CETI interact in any meaningful way, even if they are close in terms of interstellar differences.

In a basic sense, interaction over long time frames can be defined in many ways, but in terms of the Drake equation, we will define the effects of interaction (or lack of effects) by the statistical dependence between the average communicating lifetime $L$  based on the number of coexisting civilizations $N$. Therefore, if $I_{\Delta t}(N,L)$ is defined as the mutual information between the number of co-existing civilizations and their average lifetime over a measured period $\Delta t$, a basic quantification of the presence of interaction effects on the Drake Equation is

\begin{equation}
I_{\Delta t}(N,L) > 0
\label{interact-ML}
\end{equation}

The mutual information is chosen as a metric given it accounts for all types of statistical dependence, linear and nonlinear, and thus is a comprehensive description. If it is zero, it does not rule out interaction but shows that such interaction has almost no impact on the trajectory of the lifetime of interstellar civilizations. However, in the case most or all of the statistical dependence between $N$ and $L$ is linear, we can instead define the presence of interaction as the covariance.

\begin{equation}
\mbox{Cov}_{\Delta t}(N,L) \neq 0
\label{interact-cov}
\end{equation}

The covariance interpretation has the added benefit of being positive or negative depending on the beneficial or malign effects on the lifetime of civilizations based on the presence of others. While difficult to answer in the abstract, especially with sparse known data about other life in the cosmos, the answer to this question would be one of the most important regarding the dynamics of advanced intelligent life in our universe.

\subsection{A simple model of CETI interaction}

Under the conditions of the previous section where any one CETI can influence another, it is an open question how the dynamics of that influence can operate. We will assume the results of CETI interaction manifest in changes to their expected lifetime. In short, the key variables that will influence the change in lifetime are the density of CETI, given by $N$ and a measure of the strength, and direction, or their interaction given by a new variable, $r$.
The interaction parameter, $r$, so that the increase (positive values of $r$) or decrease (negative values of $r$) in the lifetime of a civilization is proportional to $r$ and $N$. For this analysis, $N$ assumes all CETI are within $L/2$ light years and can interact. Otherwise, interaction cannot be consistently expected.

Assuming an isolated CETI is expected to have a lifetime of $L_0$, then interaction effects change this to $L_0 +rN$, where $L_0$ is $L$ when $r=0$.

This relationship, where $N$ and $L$ are interdependent, can be modeled as

\begin{equation}
\begin{aligned}
\Delta N=R^{*} \times f_s \times \Delta L \\
\Delta N=R^{*} \times f_s \times rN \\
N(t)=N_0e^{(R^{*} \times f_s \times r)t}
\end{aligned}
\end{equation}

Clearly, when $r>0$, the number of CETI will increase exponential while when $r<0$, the stable point is reached when $N$ declines to zero. For both of these cases the covariances between $N$ and $L$ are positive and negative respectively.

\subsection{A more complex model of CETI interaction - the Allee Effect}

The previous model is admittedly simple and shows a simple cooperative/competitive based change in average lifetime based on the parameter $r$ and $N$. A more realistic model will bound the growth of CETI and will also account for variations in the growth rate. A common model to accomplish this is the logistic growth model and its trademark sigmoidal curve. This varies the effective growth rate based on the relationship between $N$ and a carrying capacity $K$. One drawback of the logistic model, however, is there are only two stable fixed points for the population of $N$--either $N=0$ or $N=K$. Additionally, with a logistic model, once the population is $N>0$ it will grow until it hits $N=K$. An alternate model that can account for threshold dynamics, such as a minimum number of co-extant CETI for a positive impact, and no guarantee of continuous growth in $L$ is preferable.

One longstanding model that incorporates these features is known as the Allee Effect. First discovered by W.C. Allee in observations the of how the growth rates of the flour beetle \emph{Tribolium confusum} were greatest at intermediate population densities, the Allee Effect now encompasses a wide range of effects where fitness, usually measured as the rate of population growth, varies according to the population density.

Typically, Allee Effect models have dynamics based on different regimes of growth above or below a threshold of population density. First, is minimal to negative changes in the growth rates for increasing populations at low densities. These are populations whose density or number are below the threshold $A$. Once the population density passes the threshold, $A$, it will start to show an acceleration in its growth rate until opposing forces due to population limit criteria slow it down again until the population threshold $K$ is reached where the growth rate is zero. 

The Allee effect is superior over models that only change the growth rate due to carrying capacity since it also allows the modeling of synergistic effects that can enhance growth rates in addition to carrying capacity factors that limit the growth rate. In our model, the CETI lifetime instead of the growth rate of a population will be modeled. At low densities below $A$, the lifetimes of CETIs will be similar to or lower than what is expected depending on the value of $A$ in the model. Based on the value of $A$ this models a scenario where minimal or no interaction has little effect on CETI lifetime or where CETI interaction can be negative if the threshold to synergistic positive effects is high. Once the population reaches $A$, however, the effects are modeled as mostly synergistic until the carrying capacity `K' is approached. In all models, the threshold $A$ and beyond is crucial for positive effects on the expected lifetime $L$ to take place. The model for the Allee Effect amongst CETIs here is borrowed from the model elucidated in \citep{allee,allee2}, also sometimes called the strong Allee Effect.

The Allee threshold is a number that dictates when the density of CETI will lead to mutually positive or negative effects on $L$ based on the value of $r$. If $r>0$, the case we will investigate, between $N=0$ and $N=A$, the change in $L$ is minimal negative or even strongly negative depending on the value of $A$ and $r$. However once $N \geq A$, increases in $L$ occur to due synergistic effects until they decrease again due to the effects of $N$ approaching $K$. This explicitly models a system where small numbers of co-extant CETI have little or even a negative association with each other while a greater number of co-extant CETI leads to synergistic effects within limits.


In this model $L$ changes to

\begin{equation}
L = L_0 + rN\Bigg(1-\frac{N}{K} \Bigg)\Bigg(\frac{N}{A}-1 \Bigg)
\end{equation}

Therefore, the rate of change for $N$ can be shown as

\begin{equation}
\Delta N =R^{*} \times f_s \times rN\Bigg(1-\frac{N}{K} \Bigg)\Bigg(\frac{N}{A}-1 \Bigg)
\label{allee}
\end{equation}

\begin{figure}[H]
\centering
 \begin{tabular}{cc}
 	 \includegraphics[height=2in, width=2in]{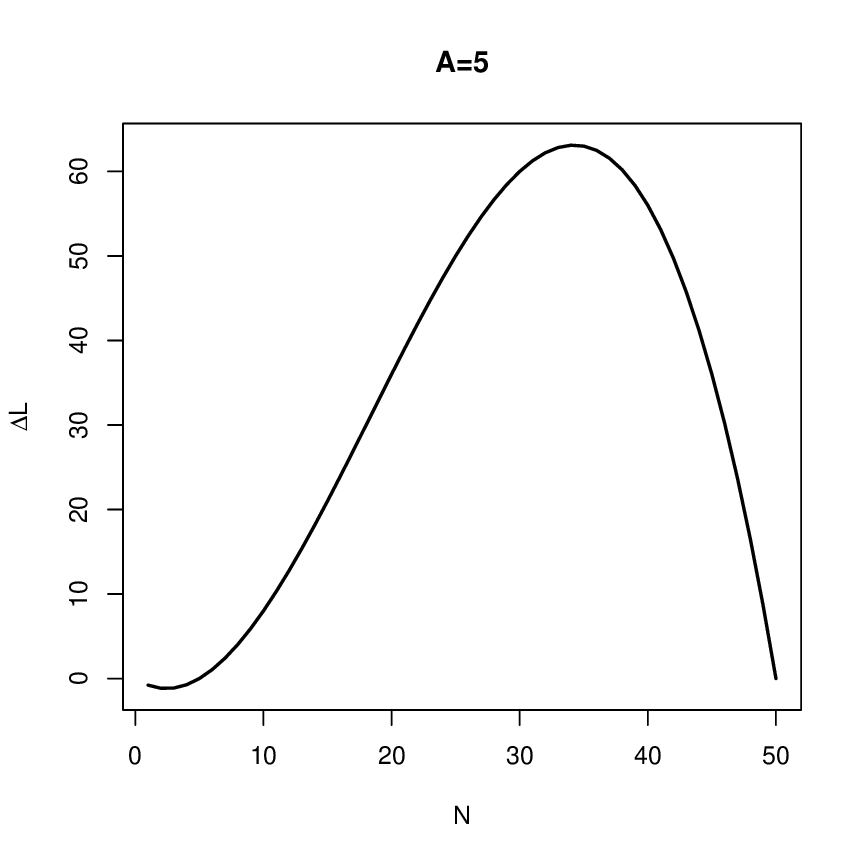} &
 	 \includegraphics[height=2in, width=2in]{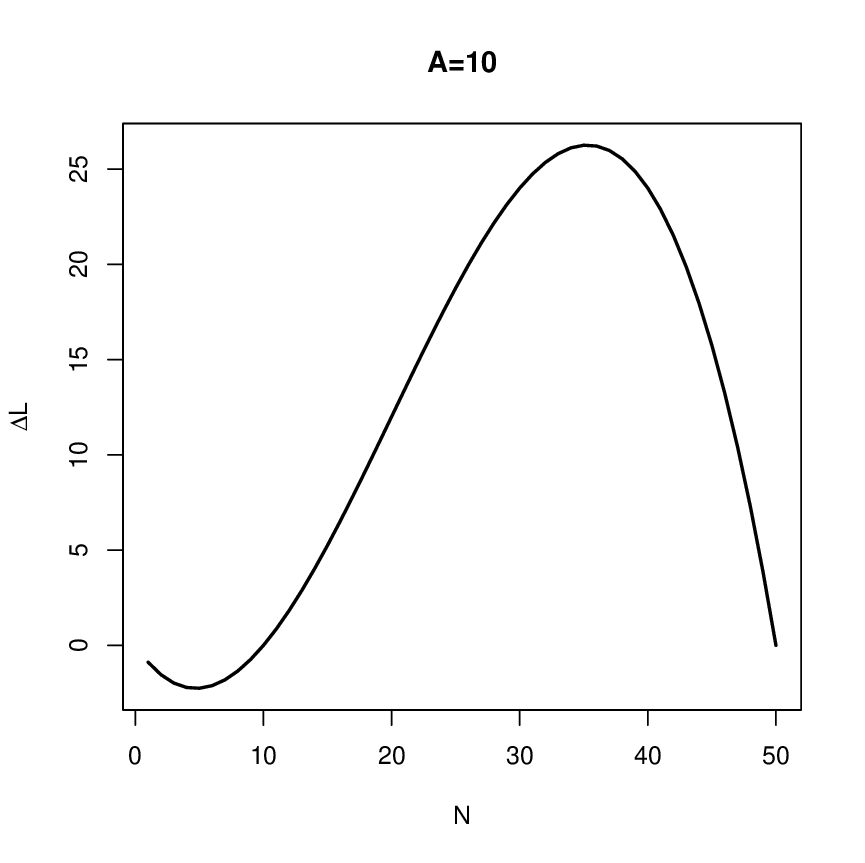} \\
     \includegraphics[height=2in, width=2in]{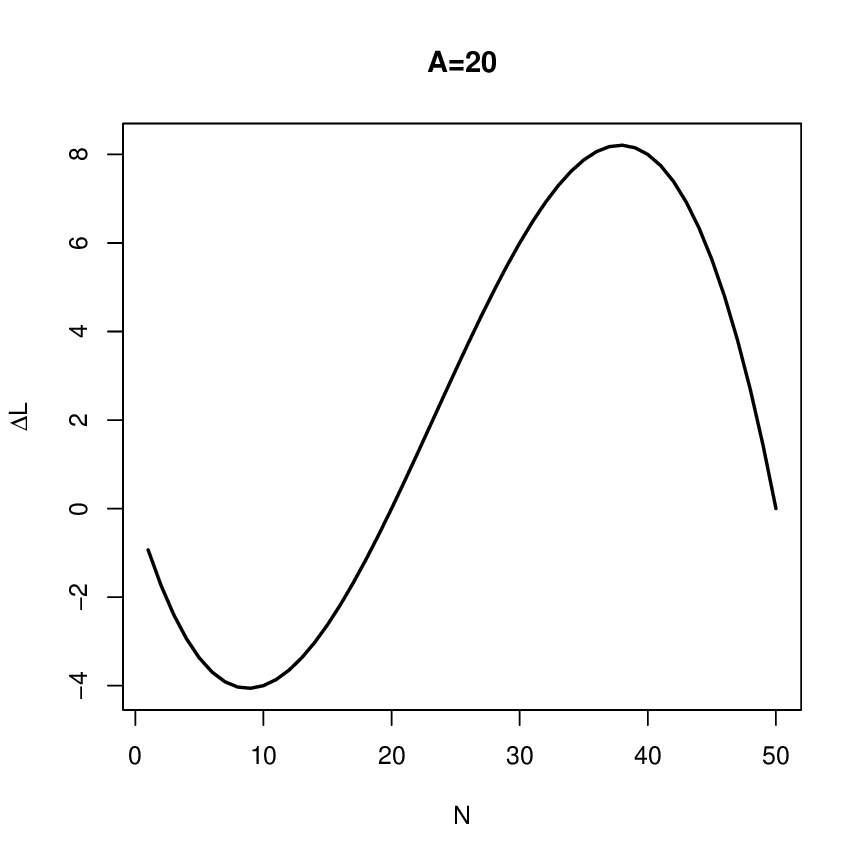}&
     \includegraphics[height=2in, width=2in]{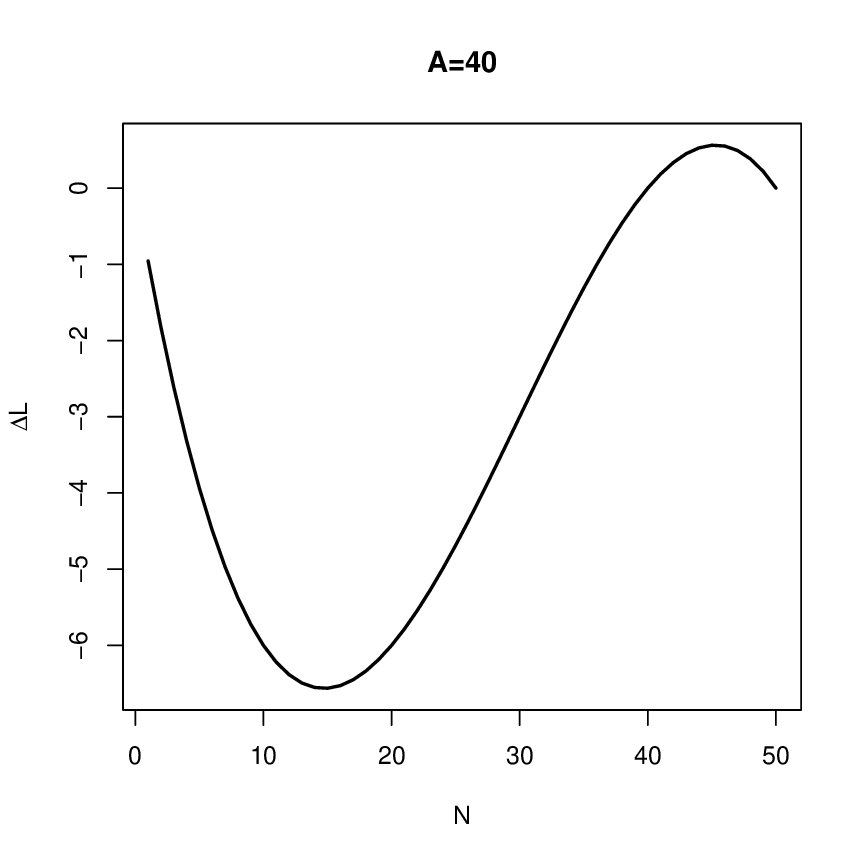}
 \end{tabular}
\caption{Plots of $\Delta L$ based on different values of $A$. Based on $L_0=1,000$, $r=1$, and $K=50$.}
\label{fig:allee1}
\end{figure}

Of note in Equation \ref{allee} there are three fixed points for $r>0$, $N \in \{0,A,K\}$ but only $N \in \{0,K\}$ are stable. At these fixed points, the formula for $N$ reduces to the traditional Drake Equation with $L=L_0$. For $N<A$, the overall rate of growth is negative leading to a declining count of CETI while at $N>A$ the growth rate increases until it hits a maximum rate and thereafter declines, still remaining positive until carrying capacity for CETIs is reached. An example of the expected behavior is given in Figure \ref{fig:allee1}.

One aspect of this expression of the Allee Effect may be unrealistic for CETI lifetimes, however. Namely, that lower densities, below the threshold $A$, would cause a decrease in the expected communicating lifetime. A more sparse universe should likely reflect the traditional Drake Equation, not represent a decline. Therefore a modified version of the Allee Effect can be used as follows

\begin{equation}
    \begin{cases}
       L=L_0, & \text{if}\ N \leq A \\
       L=L_0+rN\Bigg(1-\frac{N}{K} \Bigg)\Bigg(\frac{N}{A}-1 \Bigg), & \text{otherwise}
    \end{cases}
\label{allee2}
\end{equation}

\begin{figure}[H]
\centering
 \begin{tabular}{cc}
 	 \includegraphics[height=1.75in, width=1.75in]{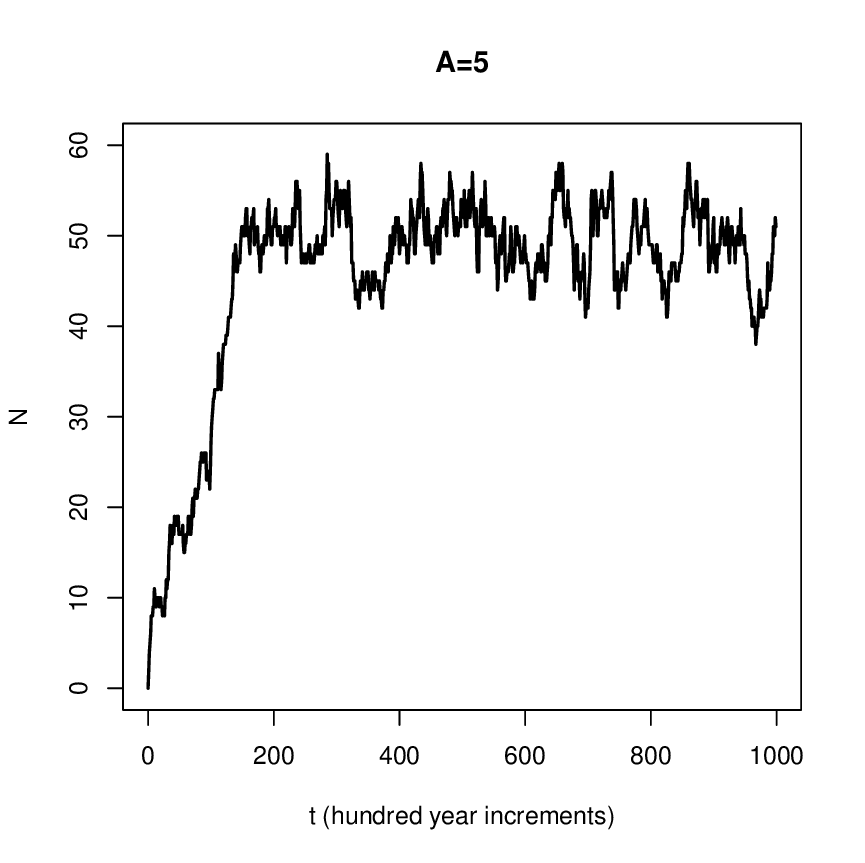} &
 	 \includegraphics[height=1.75in, width=1.75in]{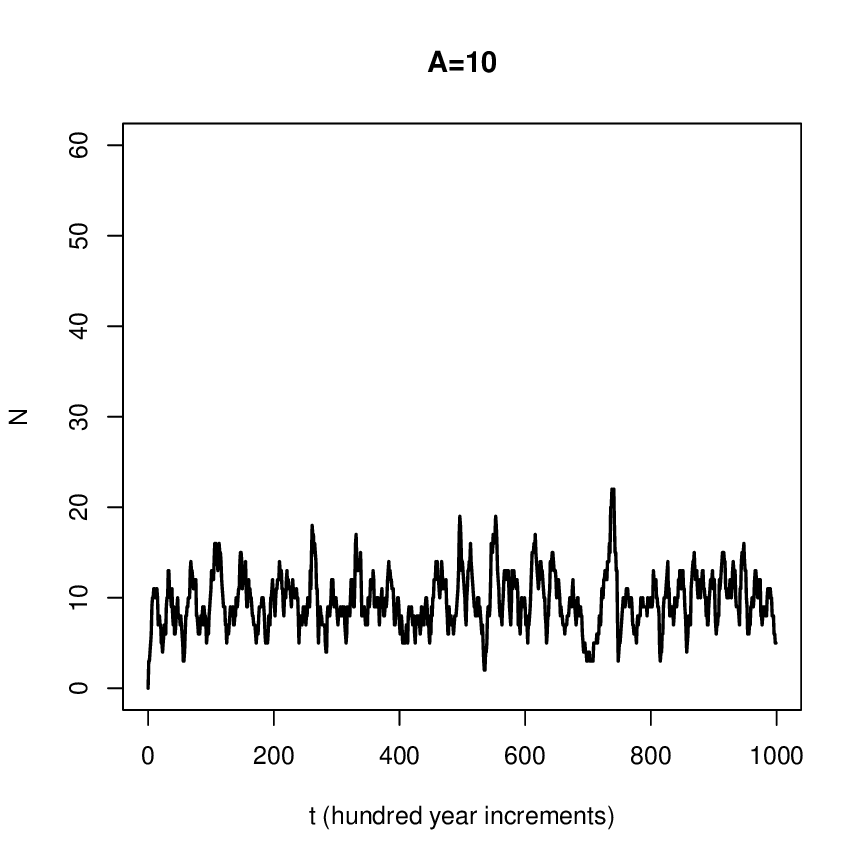} \\
     \includegraphics[height=1.75in, width=1.75in]{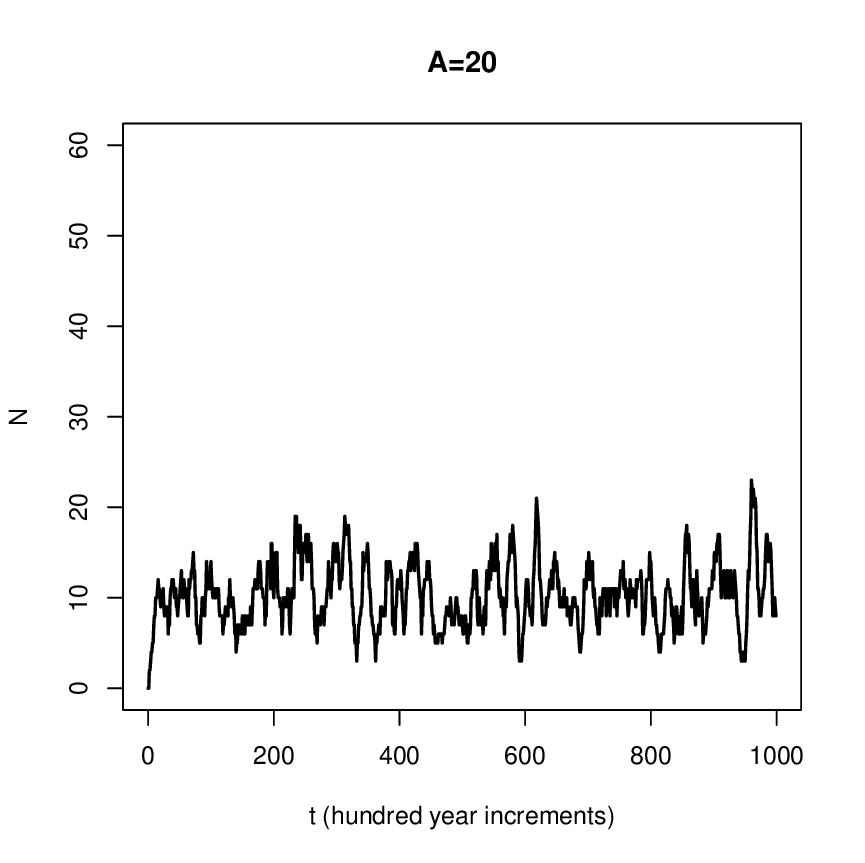}&
     \includegraphics[height=1.75in, width=1.75in]{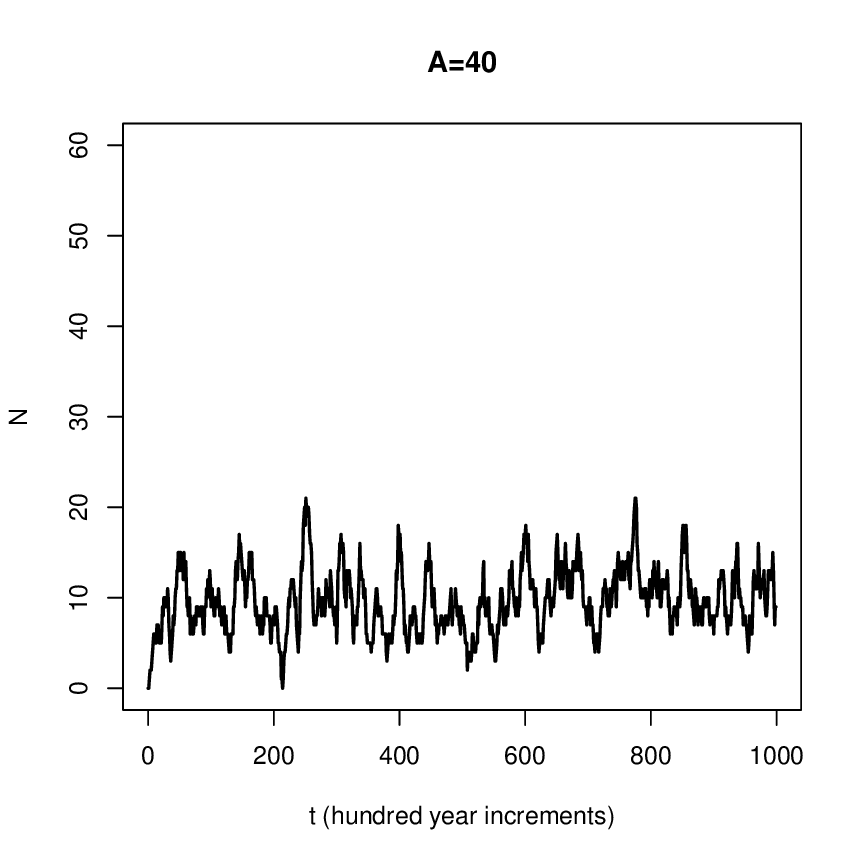}

 \end{tabular}
\caption{Plots of $N$ over time assuming CETI emergence is based on a Poisson process with $\lambda$=0.01, $L_0=1,000$, and $N_0=10$ for different values of $A$. Other model constants are the same such as $r=1$ and $K=50$.}
\label{fig:allee2}
\end{figure}

The results of the dynamics assuming Equation \ref{allee2} are shown in Figure \ref{fig:allee2}. Where $A$ is low enough for $N$ to commonly pass the threshold such as $A=5$ in this example, the number of CETI can easily exceed the expectations of the traditional Drake Equation and reach the ``carrying capacity''. Otherwise, the results basically fluctuate around the level of $N$ the traditional equation expects. Likewise, if $r<0$ (not shown), then $N=A$ is still an unstable fixed point but cannot be exceeded and the CETI population cannot exceed it for any sustained period of time and just keeps to the traditional Drake Equation expectations. In this Allee Effect model, there is no driver for sustained complete extinction of all CETI.

\section{Conclusion}

The Drake Equation continues to provide a fertile ground for research. Despite justified criticisms \citep{saganrebutb}, it remains a useful guide to theorize about the possible conditions for the presence of ETIs. Here we have used the simple assumptions of the traditional Drake Equation and a Poisson process underlying CETI emergence to build basic models of CETI interaction as well as look at relative likelihoods of communications between CETI. In short, interaction between CETI can be beneficial or detrimental with the former almost always allowing more CETI to co-exist than implied in the simple Drake Equation. On the other hand, if CETI have an overall detrimental effect on one another the Drake Equation expectations are the best case and a relatively quiet universe the worst.

The relationship between CETI was defined using changes in the average lifetime based on CETI density since it is a clear and simple way that feedback between CETI can be incorporated into the macro-trends guiding the presence of technological civilizations in space. Many other models of interaction can be postulated and developed but it is likely that all must address the question of how the overall density of CETI and their longevity are affected by their interactions.

\section{Acknowledgments}

Dedicated to Dr. Frank Drake 1930-2022.


\begin{thebibliography}{}
\bibitem[Allee et. al., 1949]{allee}\textbf{Allee, WC, Park, O, Emerson, AE, Park, T \& Schmidt, KP} (1949) \emph{Principles of animal ecology.} Philadelphia: Saunders.
\bibitem[Cirkovic, 2004]{saganrebutb}\textbf{Cirkovic, MM} (2004) The Temporal Aspect of the Drake Equation and SETI, \emph{Astrobiology} \textbf{4}, 225-231
\bibitem[Courchamp et. al., 1999]{allee2}\textbf{Courchamp, F, Clutton-Brock, T \& Grenfell, B} (1999) Inverse density dependence and the Allee effect. \emph{Trends in Ecology \& Evolution} \textbf{14}, 405-410.
\bibitem[Drake, 1961]{drake}\textbf{Drake, FD} (1961), \emph{Discussion at Space Science Board, National Academy of Sciences Conference on Extraterrestrial Intelligent Life}
\bibitem[Forgan, 2009]{forgan1}\textbf{Forgan, DH} (2009) A numerical testbed for hypotheses of extraterrestrial life and intelligence. \emph{International Journal of Astrobiology} \textbf{8}, 121-131.
\bibitem[Forgan \& Rice, 2010]{forgan2}\textbf{Forgan, DH \& Rice, K} (2010) Numerical testing of the Rare Earth Hypothesis using Monte Carlo realization techniques. \emph{International Journal of Astrobiology} \textbf{9}, 73-80.
\bibitem[Glade et. al., 2012]{poisson}\textbf{Glade, N, Ballet, P \& Bastien, O} (2012), A stochastic process approach of the Drake equation parameters, \emph{International Journal of Astrobiology} \textbf{11}, 103-108
\bibitem[Hair, 2011]{hair}\textbf{Hair, TW} (2011) Temporal dispersion of the emergence of intelligence: an inter-arrival time analysis. \emph{International Journal of Astrobiology} \textbf{10}, 131-135.
\bibitem[Kendall \& Moran, 1963]{kendall-moran}\textbf{Kendall, MG \& Moran, PAP} (1963) \emph{Geometric probability}, Griffin, 41-42.
\bibitem[Little, 1961]{little}\textbf{Little, JDC} (1961) A Proof for the queuing formula: L = $\lambda$W, \emph{Operational Research} \textbf{9}, 383–387.
\bibitem[Maccone, 2012]{statdrake1}\textbf{Maccone, C} (2012). The statistical Drake equation. In:\emph{Mathematical SETI}, Springer, 3-72
\bibitem[Prantzos, 2013]{prantzos1}\textbf{Prantzos, N}, (2013) A joint analysis of the Drake equation and the Fermi paradox, \emph{International Journal of Astrobiology} \textbf{12}, 246-253
\bibitem[Prantzos, 2020]{prantzos2}\textbf{Prantzos, N} (2020) A probabilistic analysis of the Fermi paradox in terms of the Drake formula: the role of the L factor, \emph{Monthly Notices of the Royal Astronomical Society} \textbf{493}, 3464-3472
\bibitem[Smith, 2009]{smith}\textbf{Smith, RD} (2009) Broadcasting but not receiving: density dependence considerations for SETI signals, \emph{International Journal of Astrobiology} \textbf{8}, 101-105


\end{thebibliography}
\end{document}